\documentclass{iopart}
\usepackage{graphicx,iopams,color}

\begin{document}

\newcommand{\rev}[1] {\textcolor{black}{#1}}
\newcommand{\erl}[1] {\textcolor{black}{#1}}
\newcommand{\mre}[1] {\textcolor{black}{#1}}

\newcommand{\mgb}{MgB$_2$}
\newcommand{\Tc}{T_{\rm c}}
\newcommand{\Hcii}{H_{\rm c2}}
\newcommand{\Hac}{H_{\rm AC}}
\newcommand{\Hdc}{H_{\rm DC}}
\newcommand{\degrees}{^{\circ}}

\newcommand{\fms}{f_{\rm MS}}
\newcommand{\Ims}{I_{\rm MS}}
\newcommand{\Iesi}{I_{\rm ES_1}}
\newcommand{\Iesii}{I_{\rm ES_2}}
\newcommand{\fes}{f_{\rm ES}}
\newcommand{\fesi}{f_{\rm ES_1}}
\newcommand{\fesii}{f_{\rm ES_2}}
\newcommand{\resi}{r_{\rm ES_1}}

\newcommand{\HPerp}{H_{\rm AC} \perp H_{\rm DC}}
\newcommand{\HPar}{H_{\rm AC} \parallel H_{\rm DC}}

\title{Structural studies of metastable and equilibrium vortex lattice domains in {\mgb} }

\author{E.~R.~Louden$^1$, A.~W.~D.~Leishman$^{1,2}$, C.~Rastovski$^1$, S.~J.~Kuhn$^1$, L.~DeBeer-Schmitt$^3$, C.~D.~Dewhurst$^4$, N.~D.~Zhigadlo$^{5,6}$ and M.~R.~Eskildsen$^1$}
\address{$^1$ Department of Physics, University of Notre Dame, Notre Dame, Indiana 46656, USA}
\address{$^2$ Department of Physics, Kent State University, Kent, Ohio 44240, USA}
\address{$^3$ Large Scale Structures Group, Neutron Sciences Directorate, Oak Ridge National Laboratory, Oak Ridge, Tennessee 37831, USA}
\address{$^4$ Institut Laue-Langevin, 71 avenue des Martyrs, CS 20156, F-38042 Grenoble cedex 9, France}
\address{$^5$ Laboratory for Solid State Physics, ETH, CH-8093 Zurich, Switzerland}
\address{$^6$ Department of Chemistry and Biochemistry, University of Bern, CH-3012 Bern, Switzerland}

\date{\today}

\begin{abstract}
    The vortex lattice in {\mgb} is characterized by the presence of long-lived metastable states, which arise from cooling or heating across the equilibrium phase boundaries.
    A return to the equilibrium configuration can be achieved by inducing vortex motion.
    Here we report on small-angle neutron scattering studies of {\mgb}, focusing on the structural properties of the vortex lattice as it is gradually driven from metastable to equilibrium states by an AC magnetic field.
    Measurements were performed using initial metastable states obtained either by cooling or heating across the equilibrium phase transition.
    In all cases, the longitudinal correlation length remains constant and comparable to the sample thickness.
    Correspondingly, the vortex lattice may be considered as a system of straight rods, where the formation and growth of equilibrium state domains only occurs in the two-dimensional plane perpendicular to the applied field direction.
    Spatially resolved raster scans of the sample were performed with apertures as small as 80~$\mu$m, corresponding to only $1.2 \times 10^6$ vortices for an applied field of $0.5$~T.
    These revealed spatial variations in the metastable and equilibrium vortex lattice populations, but individual domains were not directly resolved.
    A statistical analysis of the data indicates an upper limit on the average domain size of approximately \mre{50}~$\mu$m.
\end{abstract}

\maketitle

\section{Introduction}		
Vortices in type-II superconductors are of great interest, both from a fundamental perspective and as a limiting factor for applications where vortex motion leads to dissipation.
Broadly speaking, vortex matter exhibits similarities with a wide range of other interesting physical systems including skyrmions~\cite{Pfleiderer_SkL_rev,Nagaosa:2013cc},
glasses~\cite{Blatter:1994gz,Giamarchi:1995tq},
and soft matter systems such as liquid crystals, colloids, and granular materials~\cite{Nagel:2017fe}.
Correspondingly, vortex matter presents a simple model system to examine important fundamental problems such as structure formation and transformation at the mesoscopic scale, metastable states, and non-equilibrium dynamics.

The presence of metastable non-equilibrium vortex lattice (VL) phases in superconducting {\mgb} is well established~\cite{Das:2012cf,Rastovski:2013ff}.
The equilibrium VL phase diagram for this material displays three triangular configurations, denoted F, L and I,
differing only in their orientation relative to the hexagonal crystalline axes~\cite{Das:2012cf,Hirano:2013jx}.
In the F and I phases a single global orientational order is observed, with the VL nearest neighbor direction along the ${\bf a}^*$ and ${\bf a}$ directions within the basal plane respectively.
In the intermediate L phase, the VL rotates continuously from the ${\bf a}$ to the ${\bf a}^*$ orientation, giving rise to two degenerate domain orientations.
Cooling or heating across the F-L or L-I phase boundaries leaves the VL in a metastable state (MS), as thermal excitations are insufficient to drive the system to equilibrium~\cite{Das:2012cf}.
The metastability is not due to pinning, but represents a novel kind of collective vortex behavior most likely due to the presence of VL domain boundaries~\cite{Rastovski:2013ff}.

Domain nucleation and growth governs the behavior of a wide range of physical systems, and it is natural to expect similarities between the VL and for example martensitic phase transitions~\cite{Wang:2017jw}, domain switching in ferroelectrics~\cite{Shin:2007gu} or the skyrmion lattice where field/temperature history dependent metastability has also been reported in connection with structural  transitions~\cite{Makino:2017hh,Nakajima:2017uc,Bannenberg:2017ws}.
Recently, we have studied the {\mgb} VL kinetics as it is driven from the MS to the equilibrium state (ES) by an AC magnetic field~\cite{Louden:2019bq,Louden:2019wx}.
This showed an activated behavior, where the AC field amplitude and cycle count correspond to an effective ``temperature'' and ``time'' respectively.
Moreover, the activation barrier was found to increase as the fraction of vortices in the MS is suppressed, leading to a slowing down of the nucleation and growth of ES VL domains.

\mre{In this paper, we present small-angle neutron scattering (SANS) studies of the structural properties of the {\mgb} VL throughout the transition from the MS to the ES.
These complement the kinetic measurements discussed above.
Experimental details are given in Sect.~II, describing the two types of measurements used to study domain formation parallel and perpendicular to the applied field.
Results of rocking curve measurements, to determine the VL longitudinal correlation length, and raster scans, which focus on the domain formation in the plane perpendicular to the applied field direction, are presented in Sect.~III.
The implications of our data is discussed in Sect.~IV, and a conclusion is given in Sect.~V.}

\section{Experimental details}		
\label{Sec:ExpDetails}
\begin{figure}
	\begin{center}
	\includegraphics{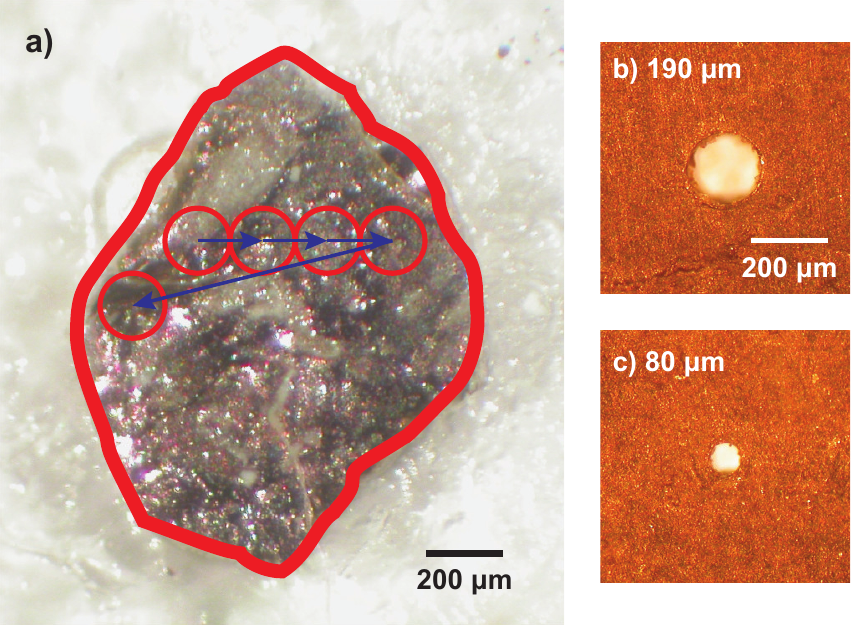}
	\end{center}
	\caption{\label{Fig:RS_Tech}
	        Schematic illustrating the SANS raster scan measurements.
			(a) An aperture is moved across the sample by translating the cryo-magnet horizontally and vertically to image the VL at each ``pixel''.
			Photos show (a) the {\mgb} crystal and (b,c) the two different gadolinium apertures used.}
\end{figure}

We used the same 200~$\mu$g single crystal of {\mgb} ($\Tc = 38$~K, $\mu_0 \Hcii = 3.1$~T) as in prior SANS studies~\cite{Das:2012cf, Rastovski:2013ff}.
The sample was grown using a high pressure cubic anvil technique that has been shown to produce good quality single crystals~\cite{Karpinski:2003fe}, and isotopically enriched $^{11}$B was used to decrease neutron absorption.
The crystal has a flat plate morphology, with an area of $\sim 1$~mm$^2$ [figure~\ref{Fig:RS_Tech}(a) and a thickness of $\sim 75~\mu$m estimated using the density of {\mgb} (2.6 g/cm$^3$). 
    
Small-angle neutron scattering measurements were performed on the CG2 General Purpose SANS beam line at the High Flux Isotope Reactor at Oak Ridge National Laboratory, and the D33 beam line at Institut Laue-Langevin~\cite{MuhlbauerRMP}.
The final data presented in this study was collected at D33~\cite{5-42-366,5-42-388,5-42-420} but consistent results were found at both facilities.
The incoming neutrons, with wavelength $\lambda = 0.7$~nm and wavelength spread $\Delta \lambda/\lambda = 10\%$, were parallel to the applied magnetic field.
Measurements were performed at either $\sim 2.5$~K or 14.2~K with 0.5~T applied parallel to the crystal ${\bf c}$~axis using a horizontal field cryo-magnet. 

Different experimental configurations were employed for rocking curves and raster scans.
For the rocking curve measurements the tightest beam collimation allowed by the D33 instrument was used, with a 10~mm diameter source aperture and an effective sample aperture of 1~mm (crystal size) separated by 12.8~m.
Combined with the effects of the wavelength spread and a VL scattering vector $q = 0.105$~nm$^{-1}$ corresponding to an applied field of 0.5~T, this yields a total experimental resolution of $0.042^{\circ}$~FWHM for the rocking curve width~\cite{Louden:2018aa}.

For the raster scans \mre{(D33 only)}, individual ``pixels'' were imaged by SANS one at a time and compiled to create a two-dimensional image of the sample, as shown in figure~\ref{Fig:RS_Tech}.
Here, gadolinium sample apertures with diameters of 190~$\mu$m and 80~$\mu$m, and a larger source aperture of 20~mm were used.
The azimuthal resolution of $4.7^{\circ}$~FWHM, estimated from the width of the undiffracted beam on the detector, was sufficient to resolve the closely spaced MS and ES VL Bragg reflections on the detector~\cite{Louden:2019bq}.
Starting with the top-left corner, the cryo-magnet was translated horizontally to image an entire row of pixels, and then moved vertically to begin the next row.
Step sizes of 200 and 100~$\mu$m for the translations were chosen to match the aperture sizes.

All VL configurations studied by SANS were prepared using the same protocol:
First, an equilibrium VL was obtained in the F phase ($T > 13.2$~K) or the L phase ($\sim 2.5$~K) 
by performing a damped oscillation of the DC magnetic field with an initial amplitude of 50~mT around the final value of 0.5~T~\cite{Das:2012cf}.
In superconductors with low pinning, this results in a well-ordered, equilibrium VL configuration~\cite{Levett:2002ba}.
Following the damped field oscillation, the ES VL was either cooled to 2.4 or 2.7~K across the F-L phase boundary to obtain a MS F phase (``supercooled") or warmed to 14.2~K to obtain a MS L phase (``superheated").
The VL relaxation is not thermal, and therefore not expected to depend on the exact oscillation temperature or the cooling rate~\cite{Louden:2019bq}.

To gradually evolve the VL from the MS to the ES phase, vortex motion was induced using a bespoke coil to apply a controlled number of AC field cycles parallel or perpendicular to the DC field used to create the VL.
A sinusoidal wave function was used, with a frequency of 250~Hz and a peak-to-peak amplitude of 0.5~mT ($\HPar$) or 7-13~mT ($\HPerp$). 
The AC field amplitudes are small compared to the damped DC field oscillation used to prepare the initial ES VL, which allowed for a precise preparation of the VL states used for the structural studies. No AC cycles were applied while the VL was imaged with SANS.

\section{Results}		

\subsection{Rocking curve measurements}
\begin{figure}
    \begin{center}
    \includegraphics{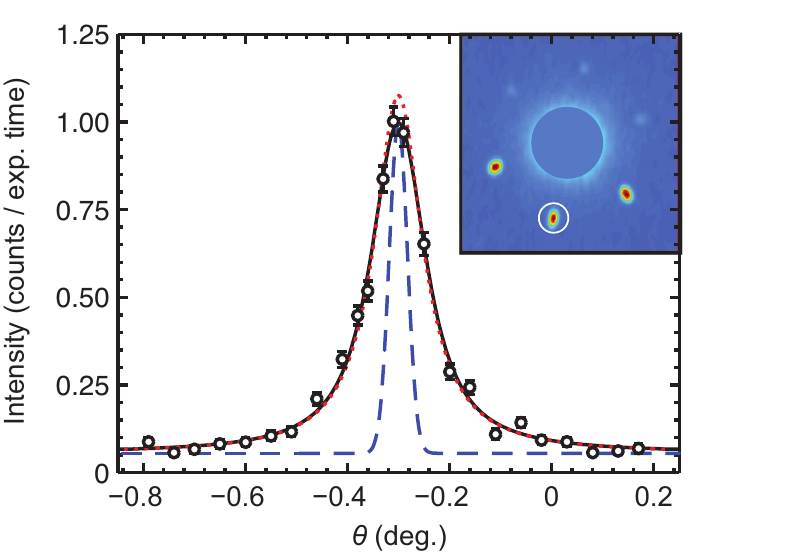}
    \end{center}
    \caption{\label{Fig:ExRC}
        Rocking curve for MS F phase VL at $T = 2.4$~K. 
	    The line is a fit to a Voigt profile, $V(\theta - \theta_0)$, given by (\ref{Eq:V}).
	    Lorentzian  and Gaussian contributions are given by respectively the dotted and dashed lines.
	    The white circle in the diffraction pattern (inset) indicates the Bragg peak used for the rocking curve.
	    Peaks at the top of the detector appears fainter as they were not fully rocked through the Bragg condition, and background scattering near the detector center is masked off.}
\end{figure}

Diffraction from the VL occurs at scattering angles given by Bragg's law: $\sin \theta_0 \approx \theta_0 = q \lambda /4 \pi$. 
As a result of both lattice imperfections and the finite experimental resolution, reflections are broadened in reciprocal space, and scattering will occur for a range of angles around $\theta = \theta_0$. 
Figure~\ref{Fig:ExRC} shows the intensity as the VL is rotated through the Bragg condition in a typical rocking curve for {\mgb}.

Spatial correlations in the VL decay exponentially with distance, with a correlation length $\zeta_L$, resulting in a Lorentzian line shape in reciprocal space.
In cases where the width of the Lorentzian and the instrumental resolution are comparable, rocking curves are best described by a Voigt profile:
\begin{equation}
    V(\theta) = \int_{-\infty}^{\infty} G(\theta') \; L(\theta - \theta') \; d\theta'.
	\label{Eq:V}
\end{equation}
This is a convolution of a Lorentzian function ($L$) representing the intrinsic width of the VL Bragg peaks and a Gaussian ($G$) representing the resolution.
The exact forms used for the Lorentzian and Gaussian functions are:
\begin{eqnarray}
	L(\theta) & = & I_0 \, \frac{w_L}{2 \pi} \frac{1}{\theta^2 + (w_L/2)^2}\\
	G(\theta) & = & \sqrt{\frac{4 \ln 2}{\pi}} \frac{1}{w_G} \exp \left[ -4 \ln 2 \left( \frac{\theta}{w_G} \right)^2 \right].
\end{eqnarray}
Here $I_0$ is the total integrated intensity of the rocking curve, and $w_L$ and $w_G$ are the full widths half maximums (FWHMs) of the Lorentzian and Gaussian.
The latter was kept constant and equal to the experimental resolution $w_G = 0.042^{\circ}$ for all fits.
The rocking curve in figure~\ref{Fig:ExRC} is almost entirely described by the Lorentzian, showing that the resolution is sufficient to allow a determination of the VL correlation along the field direction.

\begin{figure}
	\begin{center}
	\includegraphics[width = \textwidth]{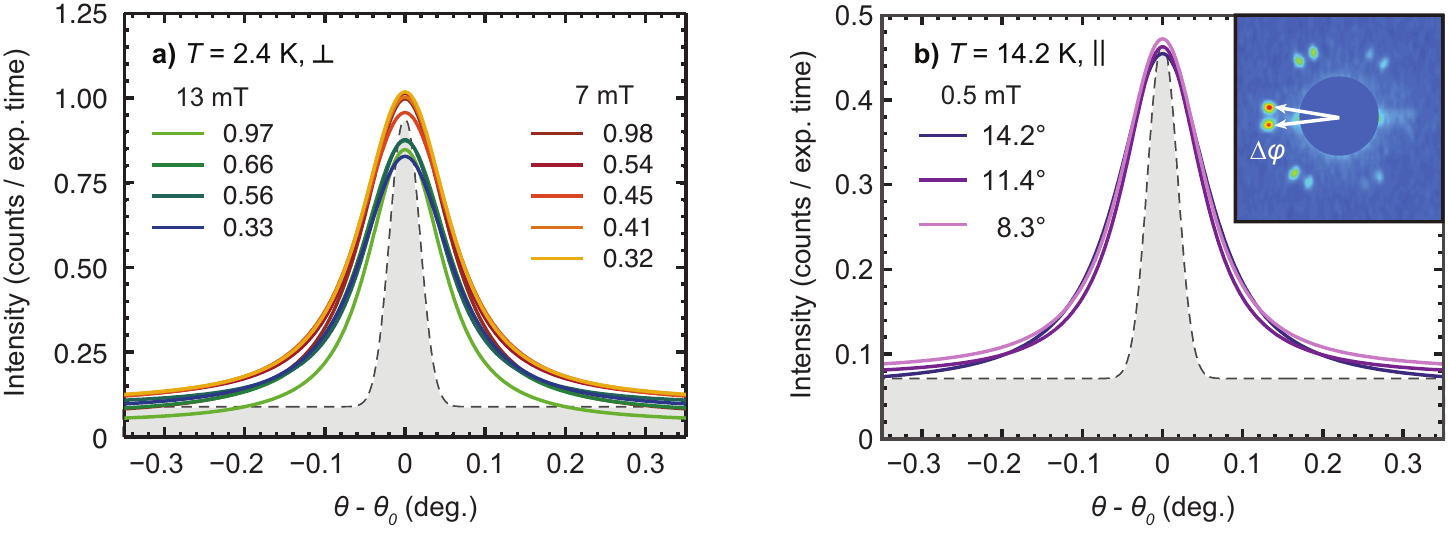}
	\end{center}
	\caption{\label{Fig:RCs}
        Rocking curve fits for measurement sequences of (a) a supercooled VL with {$\HPerp$} and (b) a superheated VL with {$\HPar$}.
        Fits are to a Voigt profile, (\ref{Eq:V}), with a fixed $w_G = 0.042^{\circ}$~FWHM indicated by the grey dashed line.
        All curves are plotted relative to their fitted center, $\theta_0$.
		The value of the transition coordinate is indicated for each rocking curve 
		by $\fms$ (a) or $\Delta \varphi$ (b).
		In (a), results for two different AC field amplitude are shown (reds/greens).
		Here, the fit for $\mu_0 \Hac = 7$~mT and
		$\fms = 0.97$ corresponds to the data shown in figure~\ref{Fig:ExRC}.
		\mre{The inset in (b) shows the azimuthal peak splitting $\Delta \phi$}.}
\end{figure}

Figure~\ref{Fig:RCs} shows rocking curves obtained at a number of configurations, as the VL is gradually driven from the MS to the ES by successive applications of AC field cycles.
The supercooled measurement sequences were carried out with $\HPerp$ (a), while the superheated sequences had $\HPar$ (b).
As seen in the following, the relative orientation of the AC and DC fields does not affect the results.
For each rocking curve the evolution of the VL towards the ES is describe by a ``transition coordinate''.
In the supercooled case, where the transition to the ES is discontinuous, this coordinate is the remnant metastable volume fraction~\cite{Louden:2019bq} discussed in more detail in Sect.~\ref{Sec:RasterScan}.
In the superheated case the MS VL domains rotating continuously towards the ES orientation~\cite{Louden:2019wx}.
Here the transition coordinate is defined as the azimuthal peak splitting, $\Delta \phi$, of the VL Bragg peaks on the detector.

In figure~\ref{Fig:RCs}, only the fitted Voigt profiles are shown for clarity.
Here, each curve is individually offset horizontally to account for small differences in the fitted Bragg center ($\theta_0$).
For the supercooled rocking curves in figure~\ref{Fig:RCs}(a), two different measurement sequences were collected with $\mu_0 \Hac = 7$~mT and 13~mT.
In the superheated case (b), a single sequence was performed with $\mu_0 \Hac = 0.5$~mT.
The order of magnitude difference of the AC field amplitudes reflects the greater efficiency of $\HPar$ in driving the VL from the MS to the ES compared to $\HPerp$.
While there are some fluctuations in the data, the widths of all of the curves in figure~\ref{Fig:RCs} are essentially indistinguishable, irrespective of preparation (superheating/cooling), the value of transition coordinate, or the AC field orientation.
The slight reduction in the scattered intensity observed for the 13~mT AC field amplitude may be due to a VL disordering in the plane perpendicular to the field direction.

\begin{figure}
	\begin{center}
	\includegraphics{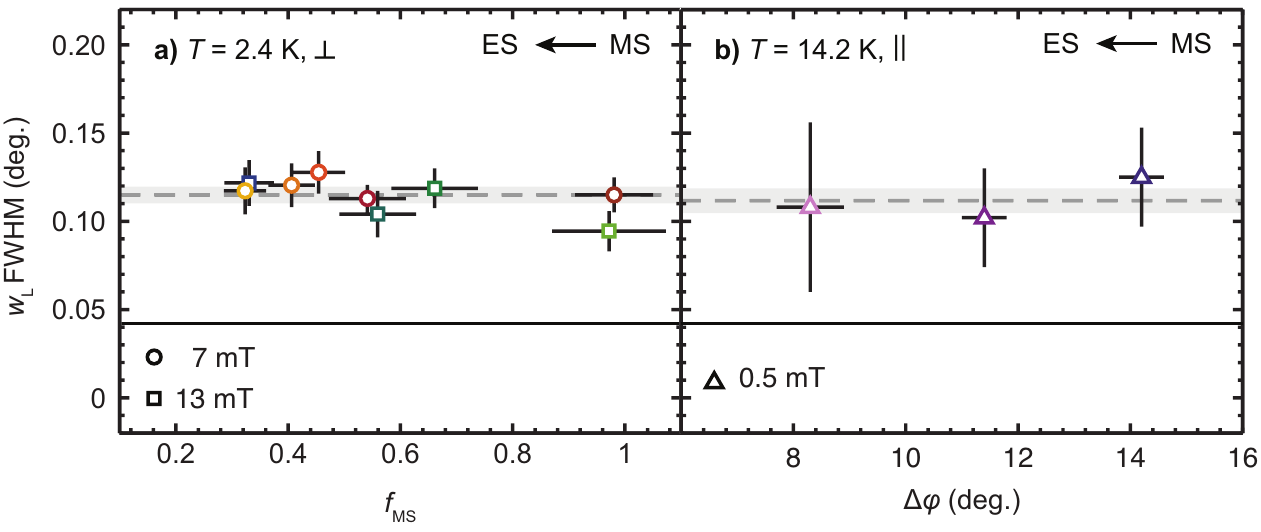}
	\end{center}
	\caption{\label{Fig:wL} 
		Loretzian widths, obtained from the Voigt fits in figure~\ref{Fig:RCs}, as a function of the transition coordinate progressing from right to left.
		For both the (a) supercooled and (b) superheated case, the dashed lines indicate the mean value of $w_L$ and the shaded areas shows the standard deviation.
		The black line shows the experimental resolution, $w_G = 0.042^{\circ}$~FWHM.}
\end{figure}

Figure~\ref{Fig:wL} shows the fitted Lorentzian widths $w_L$ as a function of the transition coordinate,
with the experimental resolution given by the solid line for reference.
For each measurement sequence, the widths are constant within the precision of the fits throughout the transition.
The average (dashed line) for the supercooled and superheated cases also agree within standard deviation for each data set (shaded area).
This shows that regardless of the transition pathway, the VL experiences no longitudinal disordering.
As previously described, the rocking curve fits are dominated by the Lorentzian contribution.
Fitting the data with a Lorentzian instead of a Voigt profile yielded widths that were at most 10\% larger than those in figure~\ref{Fig:wL}.
	
\subsection{Raster scan measurements}
\label{Sec:RasterScan}
Spatially resolved measurements of the VL were performed to investigate variations in the MS and ES domain populations in the plane perpendicular to the applied field direction.
Due to the time consuming nature of these measurements, only a single VL configuration was investigated.
Prior to the raster scans, a supercooled VL was prepared in the usual manner.
The VL was then driven to a state with approximately equal intensity in each of the three domain orientations, by applying 600 AC cycles with an amplitude $\mu_0 \Hac = 8$~mT ($\HPerp$) at the measurement temperature of 2.7~K.

\begin{figure}
	\begin{center}
	\includegraphics{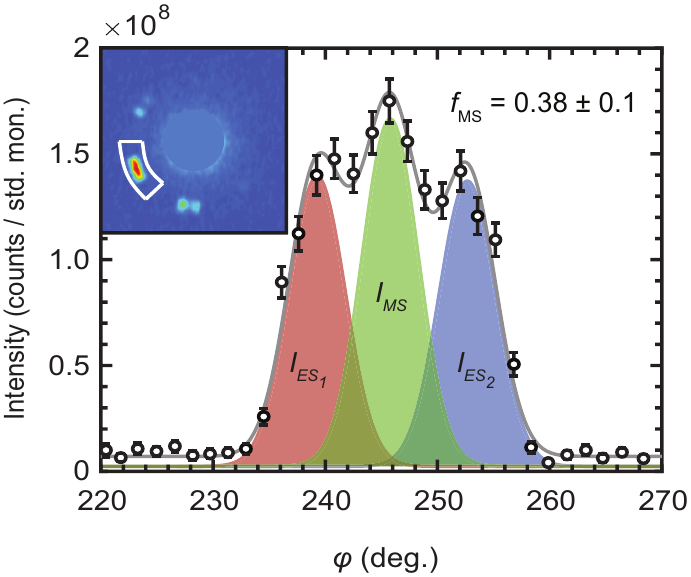}
	\end{center}
	\caption{\label{Fig:RS_state}
	Azimuthal intensity distribution using a 2~mm sample aperture to illuminate the entire {\mgb} crystal obtained at $T = 2.7$~K.
	The line is a fit to (\ref{Eq:AziI}), and shaded areas indicate the contribution from each of the VL domain orientations.
	\mre{The inset shows the detector area included in the azimuthal intensity distribution.}}
\end{figure}

Figure~\ref{Fig:RS_state} shows the azimuthal intensity distribution for the bulk system. 
The line in figure~\ref{Fig:RS_state} is a fit to a three-peak Gaussian:
\begin{equation}
	I(\varphi) = I_0 + \sum^3_{j = 1} \frac{I_j}{w_j} \exp \left[ -2 \sqrt{\log 4} \left( \frac{\varphi - \varphi_j}{w_j} \right)^2 \right].
	\label{Eq:AziI}
\end{equation}
Here $I_0$ is a constant accounting for isotropic background scattering, $I_j$ is the integrated intensity, $w_j$ is the FWHM, and $\varphi_j$ is the center for the $j$th Bragg peak.
The individual peak intensities ($I_j$) are proportional to the number of scatterers in the corresponding domain orientation.
From the fit the metastable and equilibrium volume fraction can be determined by
\begin{equation}
	\fms = \frac{\Ims}{\Iesi + \Ims + \Iesii},
	\label{Eq:fMSI}
\end{equation}
where $\Ims$ is the intensity of the central Bragg peak and $\Iesi$ and $\Iesii$ are the side peak intensities.
This yields a bulk $\fms = 0.38 \pm 0.10$ for the VL configuration used for the raster scans.

\begin{figure}
	\begin{center}
	\includegraphics{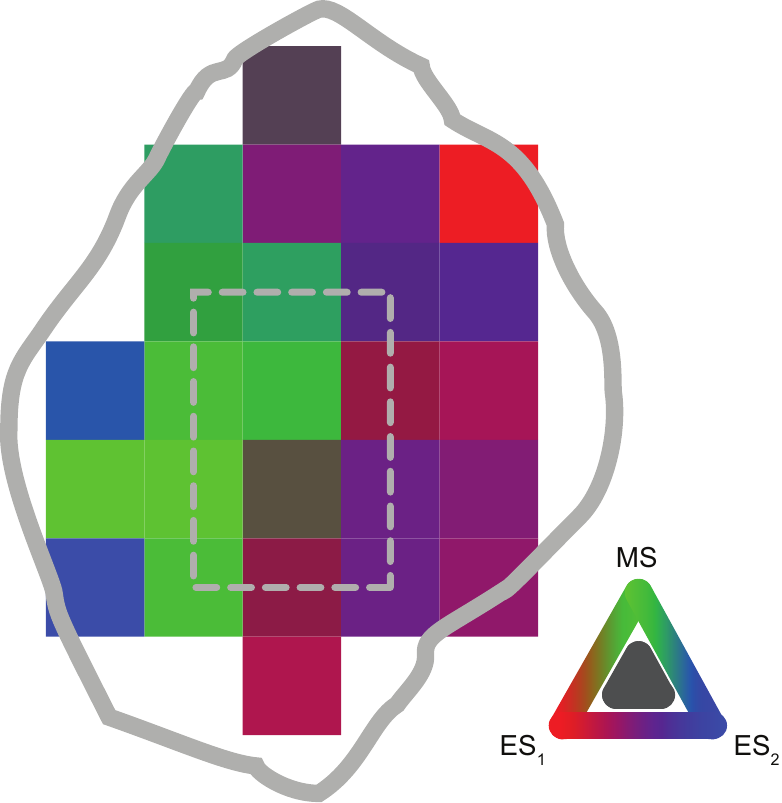}
	\end{center}
	\caption{\label{Fig:200RS}
		Raster scan collected with the 190~$\mu$m aperture at $T = 2.7$~K.
		The volume fraction of each domain orientation within each 
		pixel is indicated with an RGB color scale, 
		with red corresponding to $\fesi$, green to $\fms$, and blue to $\fesii$.
		The dashed rectangle indicates where the second, 80~$\mu$m aperture raster scan was collected on the sample.}
\end{figure}

\begin{figure}
	\begin{center}
	\includegraphics[width = \textwidth]{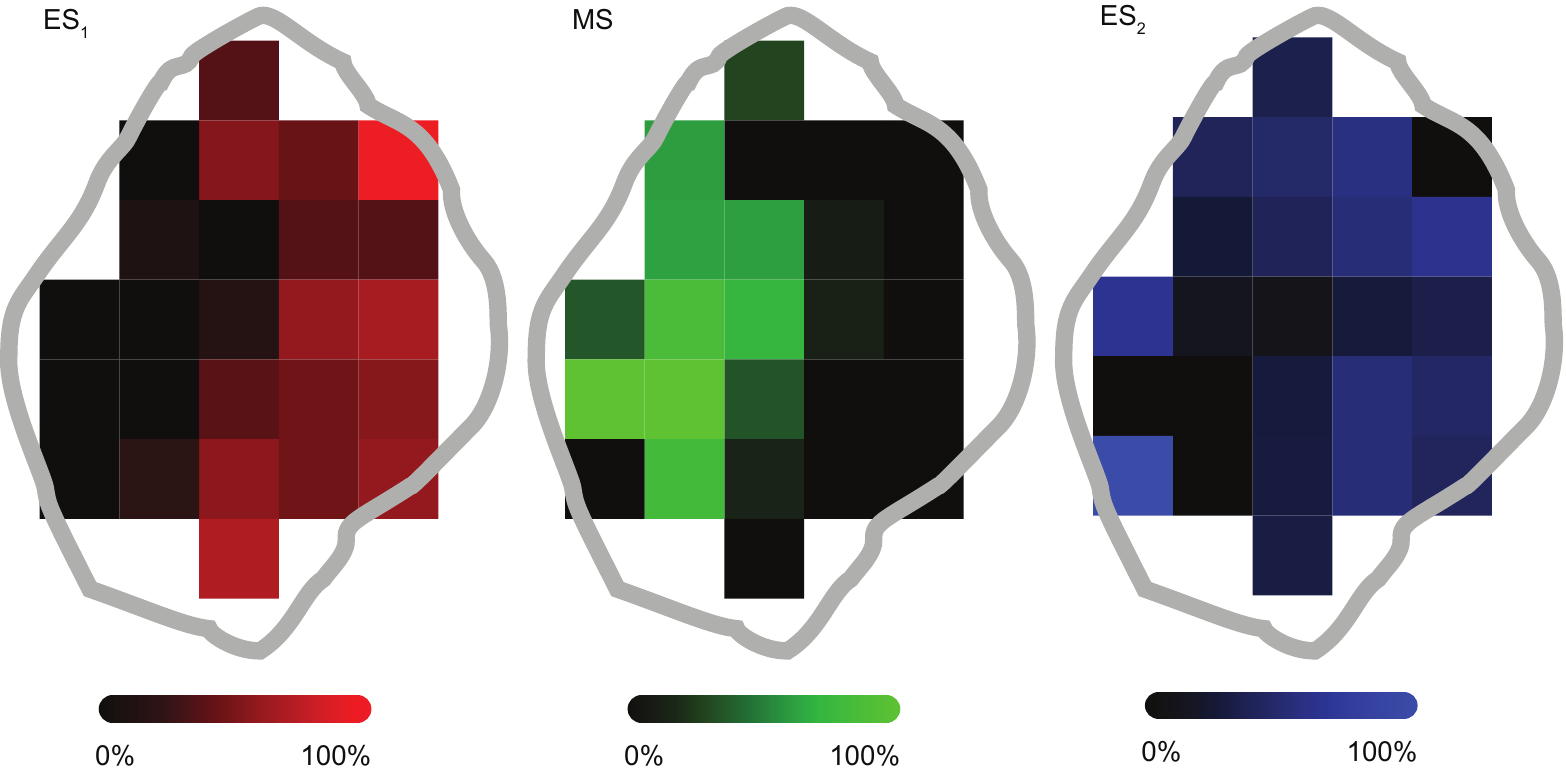}
	\end{center}
	\caption{\label{Fig:200RS_ind}
		Individual domain populations for the 190~$\mu$m aperture data in figure~\ref{Fig:200RS}.}
\end{figure}

The raster scan using the 190~$\mu$m aperture is shown in figure~\ref{Fig:200RS}.
Here the azimuthal intensity distribution for each pixel was fitted separately,
and the contribution from the three domain orientations ($\fesi, \fms, \fesii$)
was mapped onto an RGB color scale.
The equilibrium state volume fractions $\fesi$ and $\fesii$ are defined in analogy with $\fms$ in (\ref{Eq:fMSI}).
Most pixels in the raster scan contained more than one domain orientation.
The remaining MS VL domains mostly appear at the left side of the sample, and the ES domains on the right.
The two equilibrium domain orientations also were commonly found in the same pixel, see for example the purple shade in the bottom-right portion of the sample. 
Brown-grey pixels, such as at the very top and mid-third row, are a result of approximately equal contributions from all three domain orientations.
Despite being equivalent to the bulk average, such 
pixels were rare.
The domain populations are also mapped individually in figure~\ref{Fig:200RS_ind}.
	
\begin{figure}
	\begin{center}
	\includegraphics[width = \textwidth]{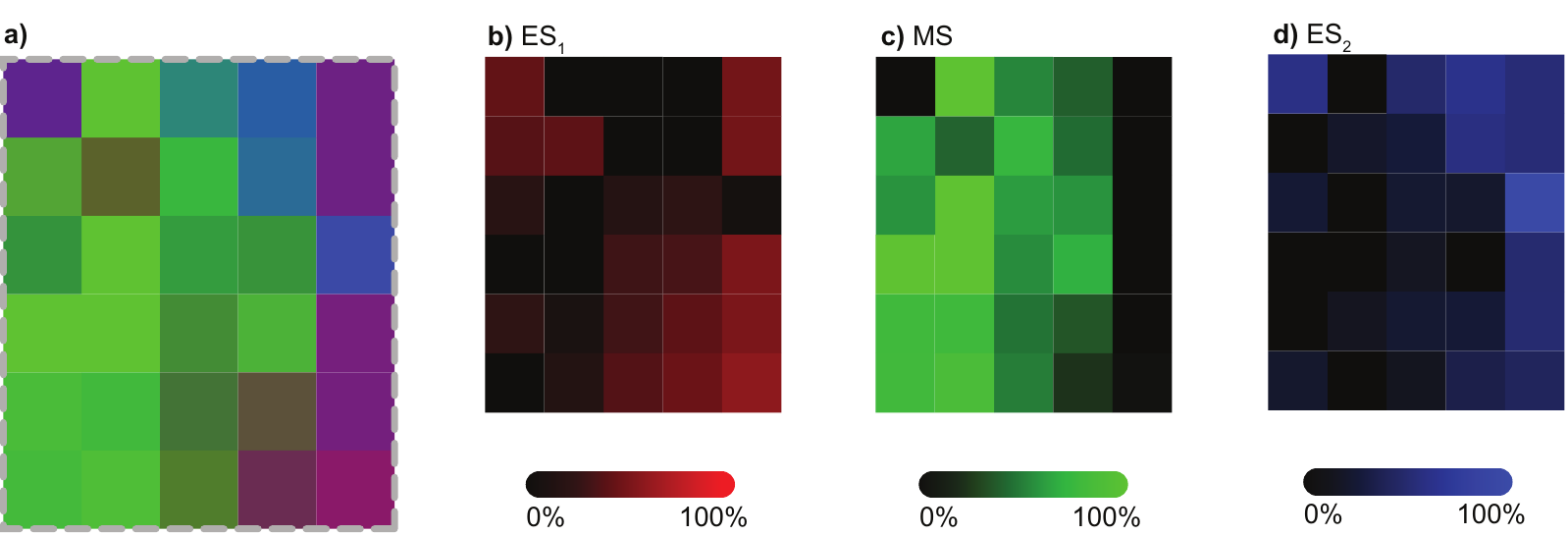}
	\end{center}
	\caption{\label{Fig:100RS}
		Raster scan collected with an 80~$\mu$m aperture at $T = 2.7$~K on the central sample region indicated in figure~\ref{Fig:200RS}.
	    (a) Volume fractions $\fesi$, $\fms$ and $\fesii$ for each pixel 
	    indicated with an RGB color scale.
		(b-d) Individual domain populations for each domain orientation.\\
		}
\end{figure}

To improve the spatial resolution a second raster scan using an 80~$\mu$m aperture was performed on the central part of the sample.
The data, visualized in the same manner as for the larger aperture scan, is shown in figure~\ref{Fig:100RS}.
Overall, the 80 and 190~$\mu$m aperture raster scans appear qualitatively similar. 
Again, pixels frequently contained a mix of domain orientations and the two equilibrium states tended to occur in the same pixel.
However, with the 80~$\mu$m aperture it is possible to discern more fine structure.
For example, there are several blue-green pixels in the top right corner indicating the presence of 
one of the equilibrium state orientations with the metastable orientation.
Similarly, towards the bottom of the scan, the pixels transform gradually from green to brown to purple.
The inability to resolve individual VL domains is unsurprising, given that the illuminated sample area with an 80~$\mu$m aperture size and an applied field of 0.5~T contains $\sim 1.2 \times 10^6$ vortices.
However, improving the resolution is not straight forward.
The present studies are already approaching the limit of the D33 SANS instrument, both in terms of intensity/required count time and the precision with which it is possible to reliably translate the cryo-magnet horizontally and vertically.
	
\begin{figure}
	\begin{center}
	\includegraphics{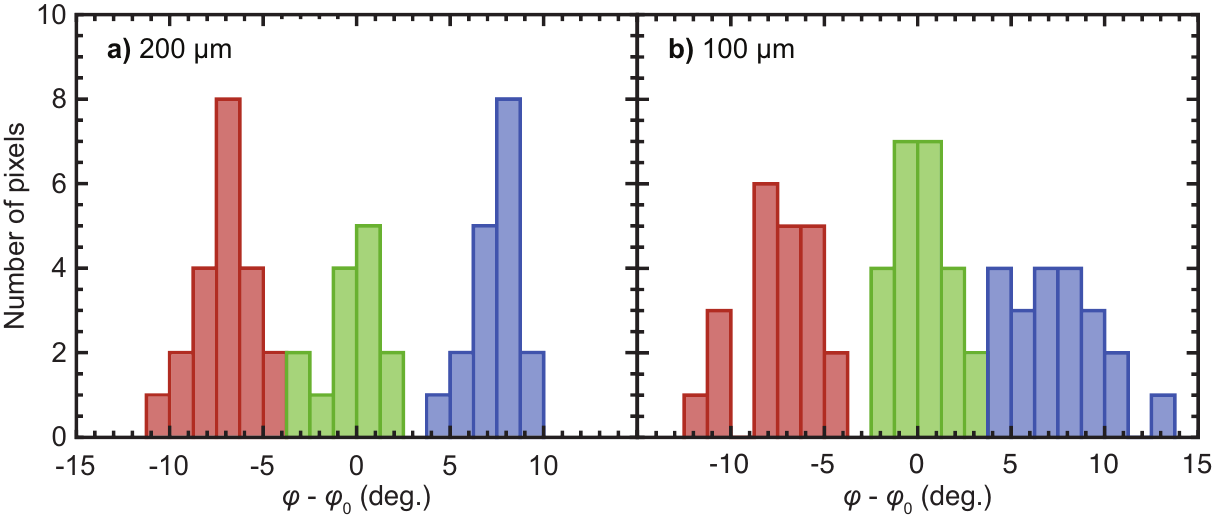}
	\end{center}
	\caption{\label{Fig:Histo_centers}
		Histogram of fitted Bragg peak centers for the three domain orientations for (a) the 190~$\mu$m and (b) the 80~$\mu$m aperture raster scans.}
\end{figure}

Compared with the bulk measurements of the VL, the $I(\varphi)$ distributions for the individual raster scan pixels exhibit a greater variation in the Bragg peak centers.
This is evident from the histograms of the fitted centers for each of the three domain orientations, shown in  figure~\ref{Fig:Histo_centers}.
Most of the Bragg peaks could be assigned a particular domain orientation (i.e. ES$_1$, MS, or ES$_2$) based on relative angular proximity to other Bragg peaks in the pixel.
For peaks that could not be inferred in this manner, the state was determined by which domain orientation angle the fitted center was closest to.

\section{Discussion}
Based on the results described above, it is possible to infer several properties of the VL phases in {\mgb} and the transition from the MS to the ES.

The VL correlation along the field direction, $\zeta_L$, is inversely related to the Lorentzian rocking widths in figure~\ref{Fig:wL}.
From the VL scattering vector $q = 0.105$~nm$^{-1}$ and the average of the mean Loretzian widths for the supercooled ($w_L = 0.119^{\circ} \pm 0.01^{\circ}$~FWHM) and superheated ($w_L = 0.111^{\circ} \pm 0.012^{\circ}$~FWHM) converted to radians, we find
\begin{equation}
	\zeta_L = \frac{2}{q w_{L}} \approx 10~\mu \mbox{m}.
\end{equation}
The correlation length is of the same order of magnitude as the crystal thickness $\sim 75~\mu$m, highlighting the high degree of ordering observed for the {\mgb} VL.
Importantly, we observe no broadening of the rocking curves and no difference between the supercooled and superheated case in figure~\ref{Fig:wL}, despite the different nature of the transition (discontinuous versus continuous).
This implies that little or no fracturing of the VL occurs along vortex direction as the system is driven from the MS to the ES,
and thus indicates that the nucleation and growth of ES state domains primarily takes place in the two-dimensional plane perpendicular to the applied field.

In principle it is also possible to infer an in-plane correlation length from the width of the VL Bragg peaks in the plane of the detector.
However, due the two orders of magnitude poorer azimuthal resolution (see Sect.~\ref{Sec:ExpDetails}) this yields $\zeta_A = 2/q w_A \sim 0.1$~$\mu$m~\cite{Das:2012cf}.
which should only be taken as a lower limit on the domain size.
Using a transverse field scattering geometry, which could take advantage of the higher longitudinal resolution of the SANS instrument to probe the in-plane domain formation, is not practical due to the plate-like morphology of the {\mgb} single crystals.

An approximate upper limit on the in-plane domain size of 80~$\mu$m is obtained from the raster scan in figure~\ref{Fig:100RS},
as almost every pixel contained scattering from more than one domain orientation. 
An estimate of the average ES domain size can also be obtained from a statistical analysis of the intensity ratio associated with the two equilibrium VL domain populations
\begin{equation}
    \resi = \frac{\Iesi}{\Iesi + \Iesii},
\end{equation}
evaluated separately for each pixel.
Assuming that the rotation of a particular MS F phase domain to either the counterclockwise (ES$_1$) or clockwise (ES$_2$) L phase orientation is equally likely,
one expects that the values of $\resi$ to follow a normal distribution.
In this case the variance is given by $\sigma_{\resi}^2 = 1/4N$, where $N$ is the number of domains within a pixel.
From the 26 pixels in the 80~$\mu$m aperture raster scan in figure~\ref{Fig:100RS} which were not purely in the MS phase
we obtain a value of $\sigma_{\resi}^2 = 0.105$. 
This yields $N \approx 2.4$, corresponding to a domain size of the order $80~\mu$m$/\sqrt{2.4} \sim 50~\mu$m.
A similar analysis can be performed on the results of repeated preparations of an ES F phase reported previously~\cite{Louden:2019bq}.
This yields 160 domains for the entire 1~mm$^2$ sample, corresponding to an average domain size $\sim 80~\mu$m.
The good agreement between these order of magnitude estimates leads us to conclude that VL domains in the plane perpendicular to the applied field is
of the order several tens of microns.

Finally, we note the similarities between the supercooled VL discussed above and structural martensitic phase transitions.
Examples of the latter include the tetragonal-to-orthorhombic transition in cuprate superconductors~\cite{Beyers:1987vq,Chen:1987cn} or the $\alpha$-to-$\epsilon$ transition in iron~\cite{Kalantar:2005hd,Wang:2017jw}.
In both cases, the system has two equal energy pathways from the initial to the final structure,
and a final configuration consisting of a periodic twin-boundary lattice rather a single global domain which has the lowest energy. 	
Here, the interface between the initial and twinned phases provides the force necessary to stabilize the metastable twin boundary lattice,
and the orientation of the interface depends sensitively on the relative populations of the two twinned phases~\cite{Barsch:1987kk}.
In the case of the {\mgb} VL, the presence of a twin boundary lattice could explain the spatial correlations between the two equilibrium domain orientations observed in the raster scans in figure~\ref{Fig:200RS} to figure~\ref{Fig:100RS}.
With the 190~$\mu$m aperture, of the 23 pixels that were not solely in the MS phase 16 (70\%) contained both ES VL domain orientations.
Similarly, for the 80~$\mu$m aperture scan, 16 of 26 pixels (62\%) show scattering intensity associated with both equilibrium states.
No pixels purely in a ES$_1$ or ES$_2$ was observed,
highlighting the preference for the equilibrium state domains to be in close proximity to each other or to a MS domain.
Further studies that could provide real space information about the VL domain boundaries, either experimentally (e.g. by STM) or by non-equilibrium molecular dynamics simulations~\cite{Pollath:2017fz,Olszewski:2018fp}, would be a valuable complement to our SANS results and interpretation.

\section{Conclusion}
We have examined the structural properties of the vortex lattice in {\mgb} as it is driven between metastable and equilibrium configurations,
using an AC magnetic field to induce vortex motion.  
Rocking curves show a lack of broadening, demonstrating that the VL does not fracture along the applied field direction in neither the supercooled nor the superheated case.
Furthermore, the VL longitudinal correlation length is comparable to the sample thickness, and the VL can be considered a system of straight rods.
Raster scans were performed to explore the formation and growth of equilibrium state domains in the two-dimensional plane perpendicular to the applied field.
While it was not possible to resolve individual VL domains, a statistical analysis provided an estimate of the average domain size of approximately 80~$\mu$m.
Finally, strong spatial correlations between the two equilibrium domain orientations is reminiscent of the twin-boundary lattice observed in connection with martensitic phase transitions.

\ack
We are grateful to J.~Karpinski for providing the {\mgb} single crystal used for this work, and to W.~Morgenlander and J.~Archer
for assistance with the SANS experiments.
This work was supported by the U.S. Department of Energy, Office of Basic Energy Sciences, under Award No. DE-SC0005051.
A portion of this research used resources at the High Flux Isotope Reactor, a DOE Office of Science User Facility operated by the Oak Ridge National Laboratory.

\section*{References}
\bibliographystyle{unsrt}
\bibliography{MgB2Structural}

\end{document}